\documentclass[a4article,11pt,openright]{elsarticle}
\usepackage{graphicx}
\usepackage{dcolumn}
\usepackage{bm}
\usepackage{color}

\usepackage{amsbsy}
\usepackage{amsmath}
\usepackage{amssymb}
\usepackage{subfig}
\usepackage{wasysym}
\usepackage{booktabs}
\newcommand{\vect}[1]{\mathbf{#1}}

\newcommand{\be}{\begin{equation}}
\newcommand{\ee}{\end{equation}}

\newcommand{\eref}[1] {(\ref{#1})}
\newcommand{\sref}[1] {Sect.\,\ref{#1}}

\usepackage[latin1]{inputenc}
\usepackage{exscale}
\usepackage{bm}
\begin{document}

\bibliographystyle{prsty}

\title{Torsion pendulum revisited}

\author{Massimo Bassan$^{1}$, Fabrizio De Marchi$^1$\footnote[4]{Corresponding author:  fabrizio.demarchi@roma2.infn.it}
, Lorenzo Marconi$^{2}$, Giuseppe Pucacco$^{1}$, Ruggero Stanga$^{2}$, Massimo Visco$^{3}$}

\address{$^1$ Dipartimento di Fisica, Universit\`a di Roma Tor Vergata and INFN, Sezione di Roma Tor Vergata, I-00133 Roma}

\address{$^2$ Dipartimento di Fisica ed Astronomia, Universit\`a degli Studi di Firenze, and INFN, Sezione di Firenze}

\address{$^3$ {IAPS-INAF and INFN, Sezione di Roma Tor Vergata, I-00133 Roma}}

\begin{abstract}
We present an analysis of the motion of a simple torsion pendulum and we describe how, with straightforward extensions to the usual basic dynamical model,  we succeed in explaining some unexpected features we found in our data, like the modulation of the torsion mode at a higher frequency and the frequency splitting of the swinging motion. Comparison with observed values yields estimates for the misalignment angles and other parameters of the model.
\end{abstract}
\maketitle

\section{Introduction}\label{intro}

The torsion pendulum has been used and studied for over  200 years and it still is 
an extremely useful, irreplaceable tool in experimental physics, allowing,  like no other instrument, to probe the regime of motion under the influence of extremely small forces.
In particular, pendulums are employed to approach, on Earth laboratories, the free fall conditions that  Test Masses (TMs) will experience in  space missions like LISA-eLISA,  where geodesic motion is a fundamental requirement for the proper operation of the gravitational wave observatory.

One would expect such an old and exploited instrument, discussed in a very   large number of papers (see e.g. the numerous references cited in the reviews \cite{RSI} and  \cite{LowEnergy}), to be, by now, fully understood and characterized.
  {However, while analyzing  the data produced by our instrument, issues arose that lead us to question the usual simple scheme consisting of a point mass suspended on an ideal, massless, flexible fibre: namely, the swinging oscillations of the fibre + TM do not describe an ellipse (nor they lay on a plane), as expected for isotropic  restoring forces, like gravity; rather, we observe a libration of the instantaneous oscillating plane.
 Although these effects are well known in mechanics,  they were never discussed, to our knowledge,  when dealing with a torsion pendulum.
We therefore developed a model for an extended mass suspended (not necessarily along its symmetry axis) on a thin elastic beam: this helped us explain some of the ``odd" features in our data.  We first recall 
the features of our data that require deeper understanding.  We then develop our mechanical model for the torsion pendulum, that allows us to fold in imperfections in the construction and suspension of the TM, defects that can lead to 
the TM center of mass not laying on the vertical symmetry axis of the system  or on the fibre direction.
Finally, we show that this extended description can help explain most of the unexpected behaviour.}\\
Although this analysis was triggered by our particular experiment and by the  above mentioned odd features, we believe that it can be of interest and help for the vast community of researchers engaged in experiments with torsion pendulums for the measurement of very small forces or torques. Indeed, we will show that, simply by looking at the output signal in the frequency band of the higher, swinging resonances (in the Hertz region),  it is possible to extract valuable information on the symmetry quality of the torsion member, be it in terms of homogeneity, construction, accuracy in hanging it or even of the presence of non symmetric force fields. These asymmetries will eventually limit the device sensitivity also in the relevant region (mHz) where the torsion signal is monitored. Therefore, such a ``fast" and simple way to monitor, understand and correct them can also be of practical interest.
\section{Experimental set up and measurements} \label{sect2}
In preparation for the flight of LISA-Pathfinder \cite{LTP},  a space mission that will serve as a technology demonstrator for LISA, the torsion pendulums help us understand and characterize all possible sources of spurious noise that can affect the free fall of a Test Mass in geodesic motion.  To this purpose, successful pendulums have been developed in the Universities of  Trento \cite{trento1,trento2} and Washington \cite{schlamm}, where  many possible sources of unwanted perturbation have been characterized and measured: electrostatic forces, residual magnetic coupling, damping from residual gas, thermal gradients and so on.

\begin{figure}[h!]
\hspace{-1cm}
\scalebox{0.65}{
\includegraphics[width=0.9\columnwidth]{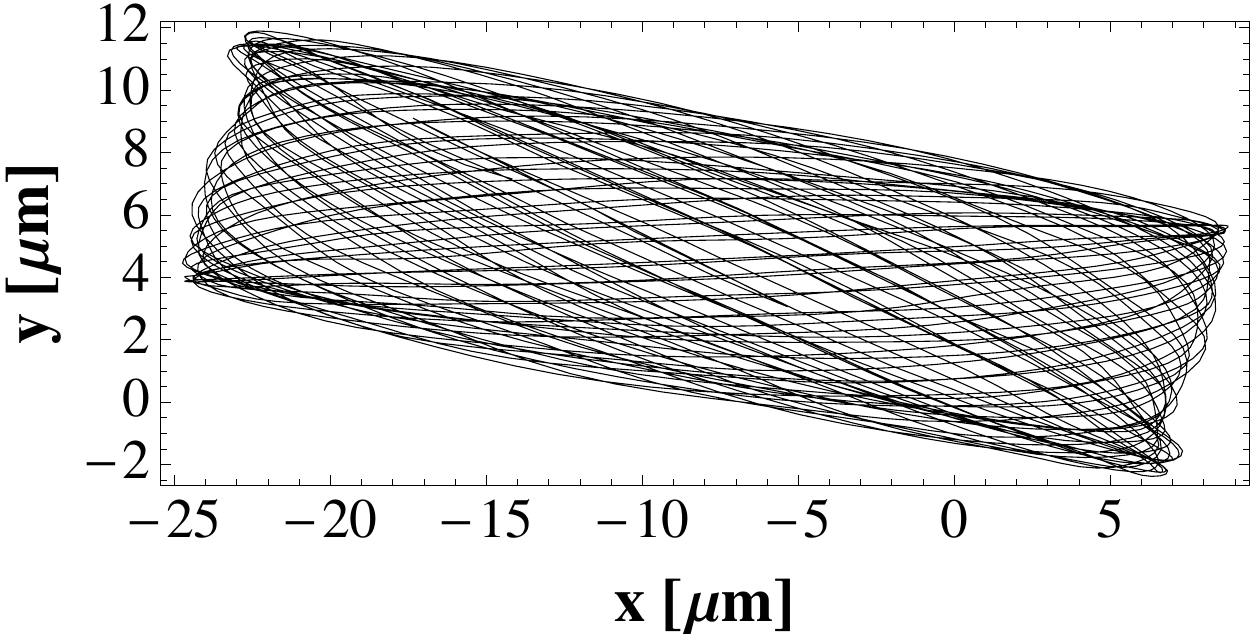}
\includegraphics[width=0.8\columnwidth]{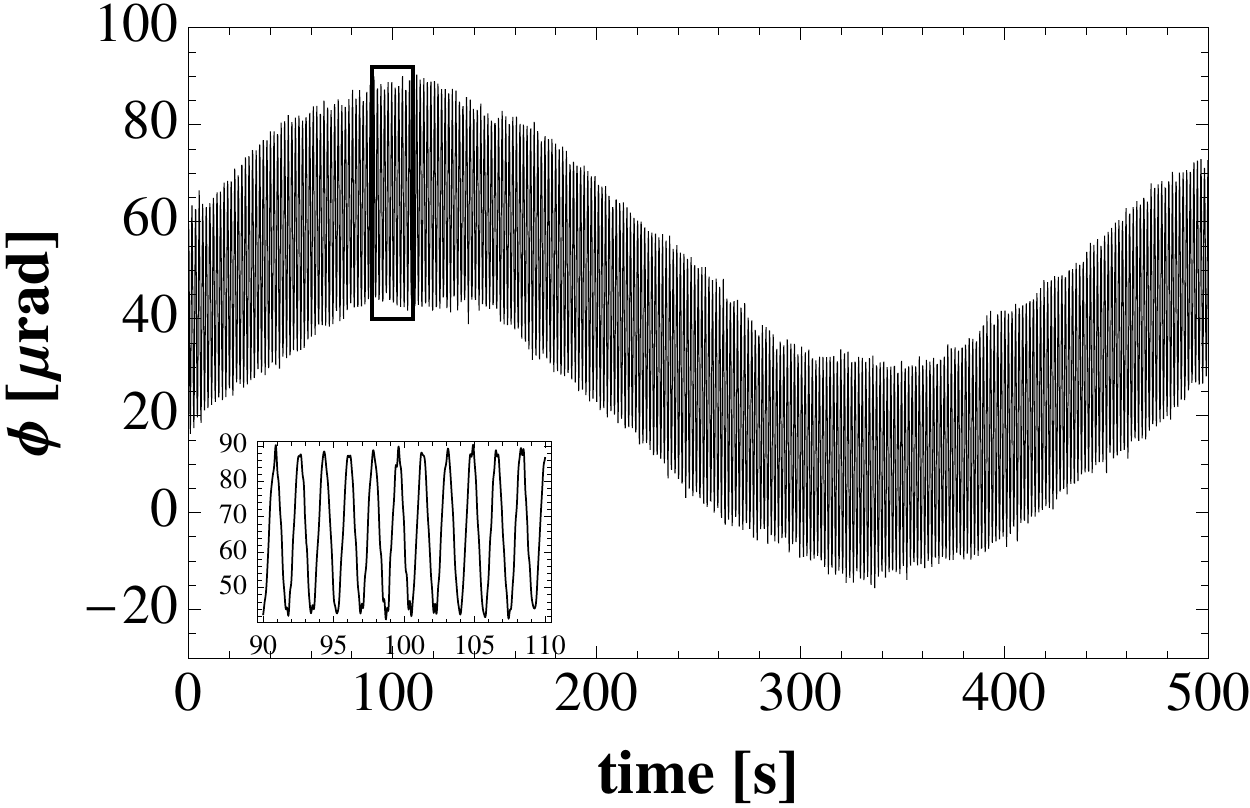}
}
\caption{Left: Lissajous figure as described by pendulum data (time span: 80s). Note that $x$ and $y$ axes have different scales. Right: torsion of the TM from data: the modulation at the swinging frequency is clearly visible in the zoom insert.}
\label{torsione_e_pendolo}
\end{figure}

\begin{figure}[h!]
\centering
\includegraphics[width=0.7\columnwidth]{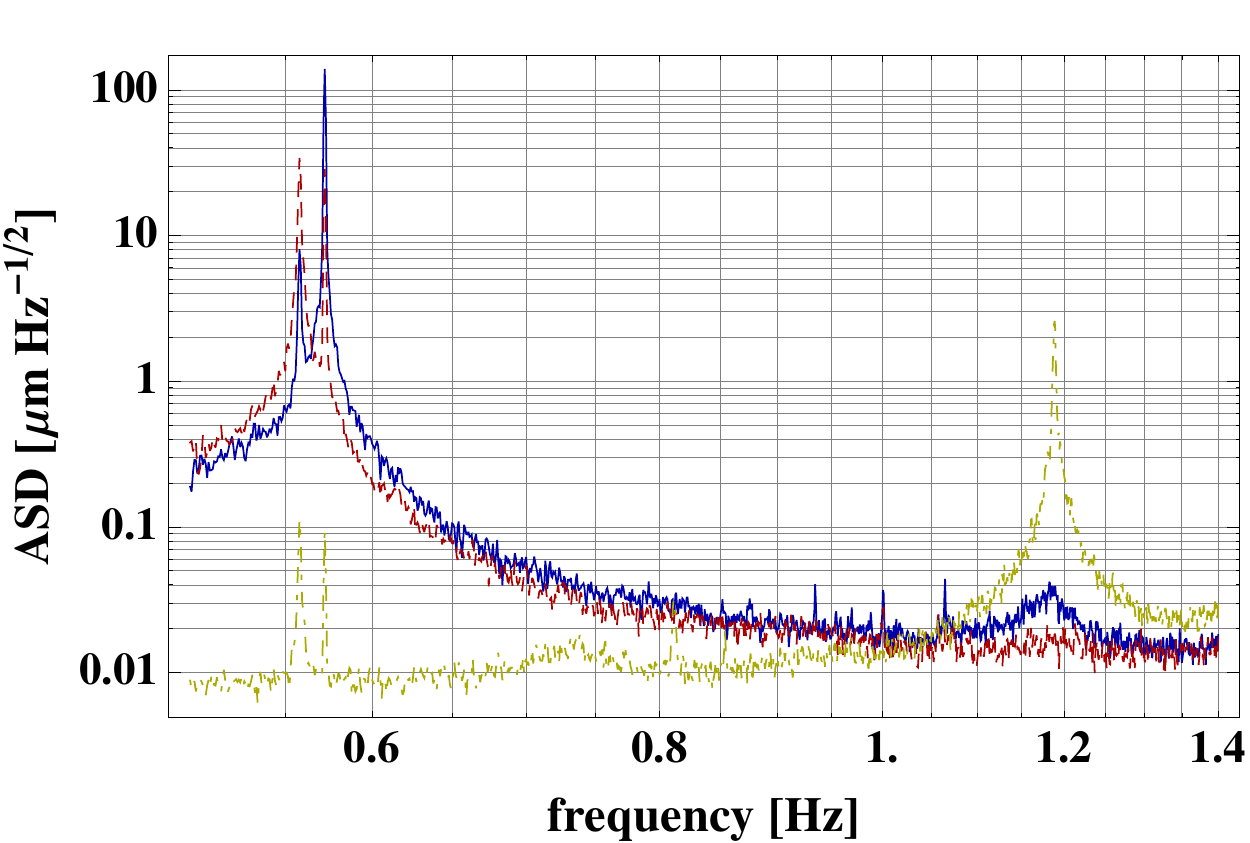}
\caption{Spectra of the observed data. Swinging modes (doublet at  $~0.56$ Hz) in the spectra of $x$ (thick), $y$ (dashed) and $z$ (dot-dashed) and bouncing mode in $z$ at 1.2\,Hz: the latter is an alias, generated by the sampling at 10 Hz, of the 8.8\,Hz longitudinal vibration mode of the fibre.}
\label{spettri_dati1}
\end{figure}

Real life in space will be more complex: each  TM in flight will behave as a particle in free fall, i.e. free from any experimenter's intervention, along  the sensitive direction (the one that faces the opposite, far TM in the interferometer arm), while motion in all other degrees of freedom (DOFs) will be restrained by feedback, to keep both TM centered inside one spacecraft \cite{pathfinder}. 
To address this issue, we are completing a complementary instrument, nicknamed  PETER (PEndolo Traslazionale E Rotazionale, {\it Rotational and Translational Pendulum}) \cite{PETER}: a double torsion pendulum where force-free motion is to  be achieved simultaneously  in two different DOFs, i.e. for translation of the TM and for rotation around its axis.  
This is achieved via a first torsion fibre, supporting a crossbar. From the end of one of the crossbar arms, a second fibre hangs, that supports a cubic Test Mass. 
 {The TM is enclosed in a \emph{Gravitational Reference Sensor} (GRS), a hollow metal box padded with electrodes.
 This set of capacitors allows us to detect motion of the TM along all 6 DOFs, as well as to apply electrostatic biases to adjust or stabilize the TM position \cite{SI}.
In the first steps of the commissioning of this apparatus, we operated it as a traditional, single DOF torsion pendulum, by fastening the crossbar on a rigid support and keeping  the sole lower fibre in operation. The 
65 cm long, $25 \mu m$ diameter W fibre, supporting a hollow Al  cubic TM, 46 mm in side, for a total mass of $0.106$ kg results in an eigenfrequency for the torsional mode of 2.2\,mHz. The simple, swinging pendulum motion, evaluated in the approximation of a lumped mass hanging from a rigid cable, is expected at 0.56\,Hz, while the bouncing motion, due to longitudinal oscillations of the loaded fibre, takes place at a frequency of 8.8 Hz. 

The motion of the TM is monitored through linear combination of the output of the 6 impedance bridges. 
Each bridge has 2 facing capacitors on the sensitive arm: for example, the sum of the two signals $X_{left}$ (read by  $X_{left-front} - X_{left-back}$) and $ X_{right}$  yields the mean position of the TM center along X, while their difference, divided by the distance between adjacent electrodes, gives the $\phi$ angle of rotation about the $z$ (vertical) axis.  

In the data recorded}, one expects to find the torsional motion in the $\phi$ channel,  the swinging motion in $x$ and $y$ channels, and the bouncing motion in the $z$ channel. However, while analyzing the data,  we encountered a few unexpected features  that we  had to explain:
\begin{enumerate}
\item {The swinging mode is actually}
a ``doublet'' at the frequencies 0.558 Hz and 0.572 Hz. 
Either line can be suppressed, leaving just the lower in the $x$ spectrum and the higher in the $y$ spectrum (or viceversa), by applying to the $x$ and $y$ time series a rotation of about $-13$  (or  $-13 - 90$) degrees {in the  $xy$-plane.}\\
In the time domain, by plotting $y$ vs $x$,  the doublet corresponds to Lissajous figures on the $xy$-plane and the envelope of these is tilted of about $-13$ degrees {(Fig.\,\ref{torsione_e_pendolo})}.  These figures are completed in 72\,s, a time quite shorter than the torsion period of 454\,s. To separate the lines is equivalent to ``straighten'' the envelope axes.
\item The doublet is also present  in the $z$ displacement spectrum {(Fig.\,\ref{spettri_dati1})}: in the simple pendulum swinging, vertical motion is a second order effect  and should therefore appear, if at all,  at twice the doublet frequencies.\\
However, the doublet lines in $z$ can be suppressed  by rotating the data in a suitable way:  by trial and error, we find
that, by rotating the data by two small angles, first in the $xz$-plane and then in the $yz$-plane, the doublet disappears  from the $z$ spectrum.
\item The $\phi$ spectrum shows, in addition to the torsion line at 2.2 mHz, the doublet of the swinging motion: in other words, the torsional oscillation is modulated at the swinging frequencies {(see \sref{torsione_e_pendolo})}. This is also an unexpected feature  in a torsion pendulum  {(Fig.\,\ref{spettri_dati1})}.
\end{enumerate}


\section{Extending the pendulum model}
In order to explain these features, we developed a more flexible model for the torsion pendulum:  the basic feature of this is to consider a rigid body suspended to the fibre at an arbitrary point, therefore not necessarily associated to any particular symmetry of the body. This reflects the practical fact that, despite maximum experimental accuracy, some misalignment can always occurr when the fibre is fastened to the TM, so that the fibre and a symmetry axis of the TM might end up not lying exactly on the same line.

\subsection{Reference frame and configuration angles}

\begin{figure}[h!]
\centering
\includegraphics[width=0.42\columnwidth]{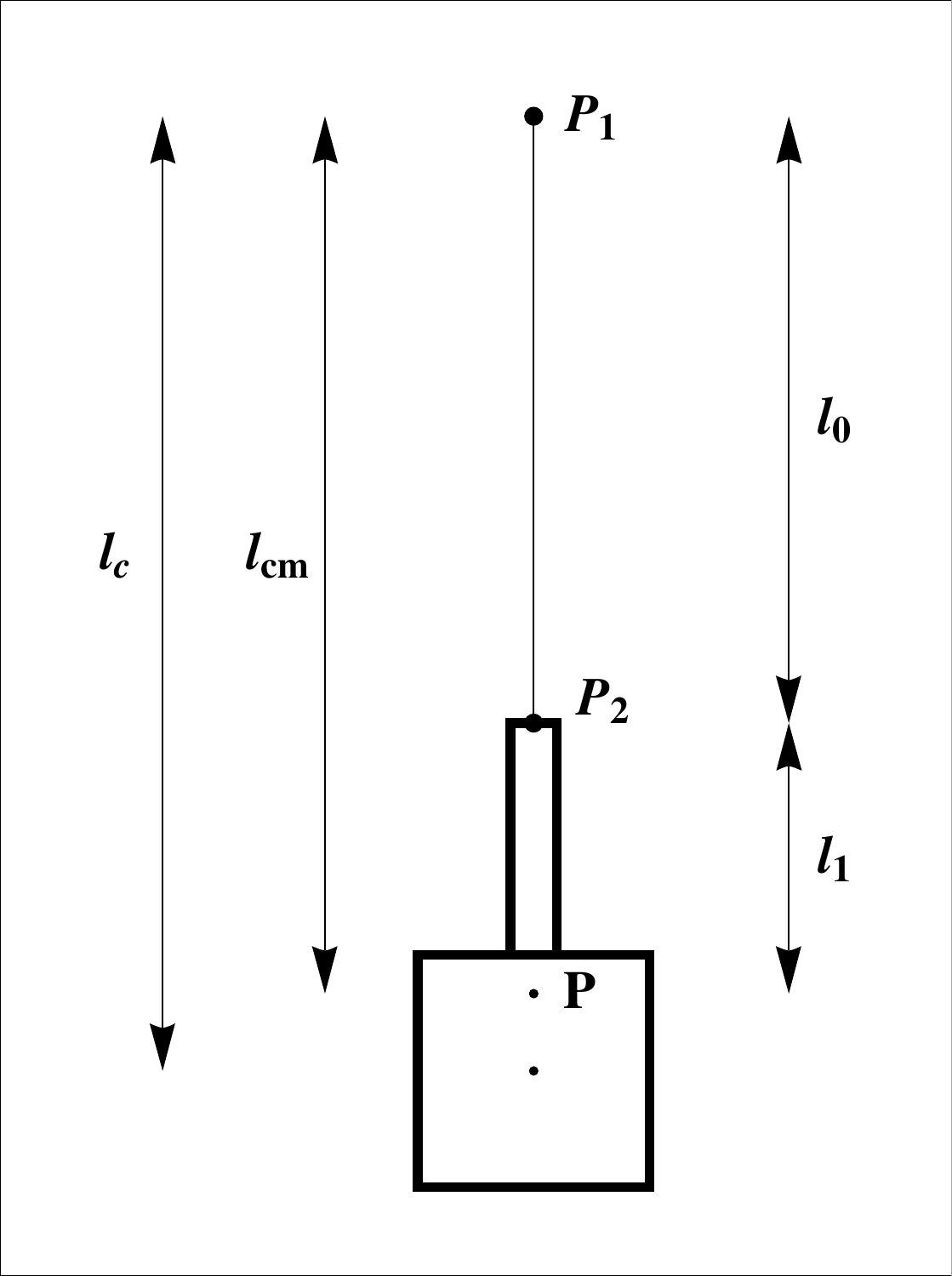}\hspace{1.5cm}\vspace{-0.1cm}
\includegraphics[width=0.38\columnwidth]{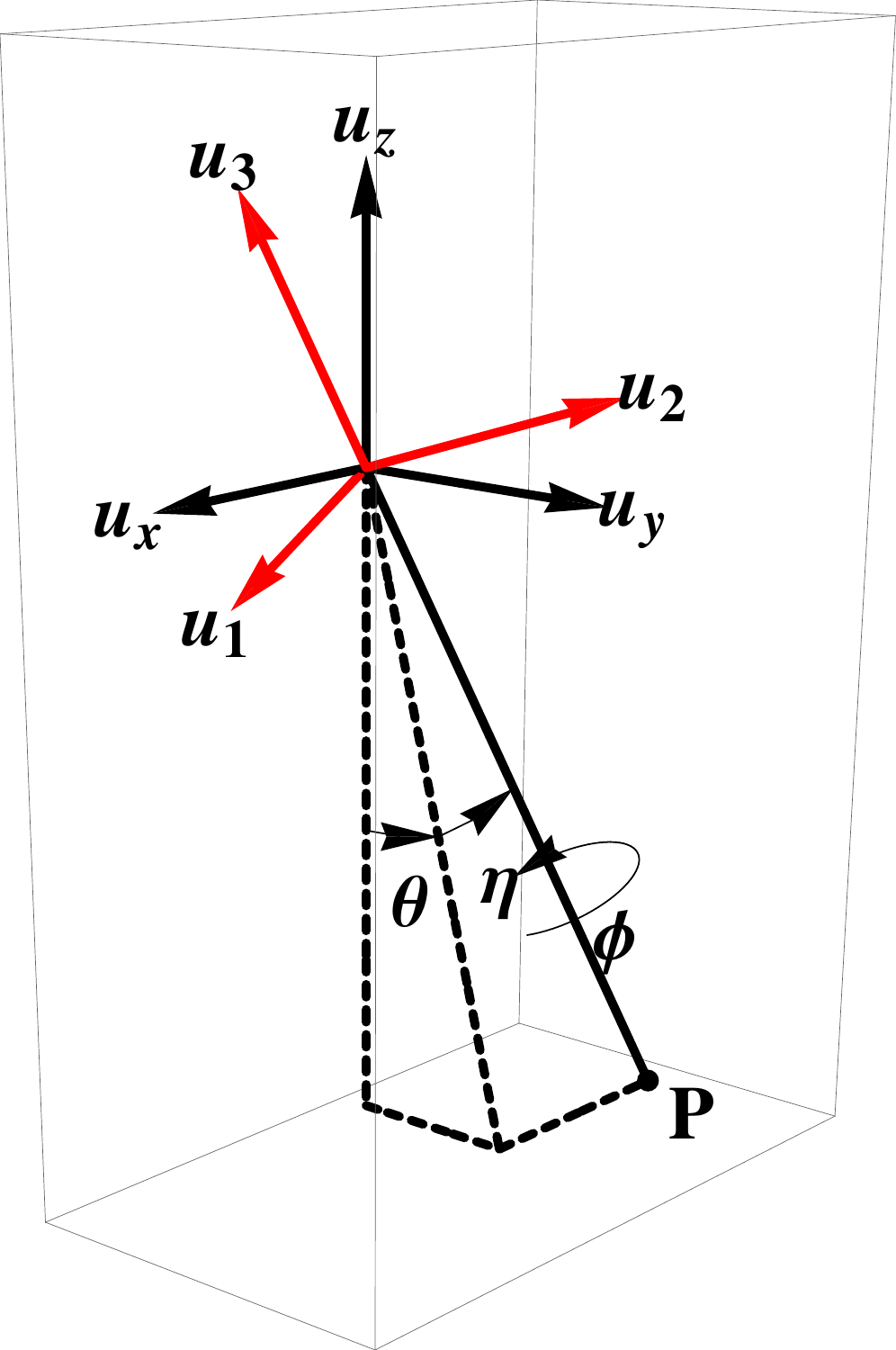}
\caption{Left: Definition of the  lengths mentioned in the text:  we neglect here both the dynamic (bouncing motion) and the  static elongation of the fibre, and therefore the difference between $\ell_0$ and $\ell_0'$ (see sect.\ref{tretre}), $P$ is the position of the center of mass. The center of the cube is also shown, at a distance $\ell_c$ from  $P_1$.  
Right: moving coordinate system: $\bm{u}_x,\bm{u}_y,\bm{u}_z$ represent the inertial frame, while $\bm{u}_1,\bm{u}_2,\bm{u}_3$ identify the coordinate system at rest with the rigid body.}
\label{angoli_cinesi}
\end{figure}

The system we want to describe consists of a thin fibre hanging from a fixed point $P_1$: at the lower end of the fibre, in the point $P_2$, at a distance $\ell_0'$ (along $z$ at rest) a solid, symmetric TM is suspended. The center of mass of the TM is at a distance $\ell_1$ below $P_2$, hopefully, but not necessarily, along  the ``vertical" principal axis of inertia of the TM  {(see fig.\,(\ref{angoli_cinesi}a))}. In our case the TM is a hollow Al cube plus a suspension shaft, but the presence of a rectangular mirror on the shaft breaks the symmetry, so that the two principal moments of inertia perpendicular to the $z$ axis are not equal.

To represent the motion of a rigid body, besides a fixed reference frame, we need a frame at rest with the body (moving frame). 
The fixed, inertial, laboratory frame, labeled  by indices  $x,y,z$, is defined by the sides of the GRS, and therefore by the outputs of the electrostatic readout. 
Its origin is set at the upper suspension point $P_1$. The $z$ axis can differ from the direction of the local gravity if we take into account ground tilt. This effect must be considered separately and will not be addressed here.
We will use indices $\{1,2,3\}$ for the axes of the moving (with the Test Mass) frame.
A suitable choice of comoving frame is such that, for small oscillations, it coincides with the fixed frame, apart a possible translation of the origin.

The most immediate choice to define the coordinate system co-moving with the rigid body would be to use polar coordinates and Euler angles. However, this is not  convenient to describe small oscillations, because, as the aperture angle tends to zero, the azimuthal angle becomes undetermined.
We adopted a more suitable coordinate system \cite{cinesi1}, \cite{cinesi2}, shown in fig.\,(\ref{angoli_cinesi}b): we position the origin of moving frame $123$ in the center of mass of the TM, with the axes  aligned with the principal axes of inertia of the TM.
 In order to align the ``3" axis of the moving frame with the instantaneous fibre direction, 
 {we perform the following rotations:  first, by an angle $\theta$ around  the $x$-axis; then by an angle $\eta$  around  the {new $y'$}-axis; finally by  $\phi$ around the $z$-axis:
 in this, we follow the convention \cite{trento1} for indicating rotations of the TM}. { Note that, in usual control (and flight dynamics) jargon, one would refer to $\phi$ as the \emph{yaw} angle, $\theta$ as the \emph{roll} angle, and 
$\eta$ as the \emph{pitch} angle.}

{Defining $\ell_{cm}=\ell_0'+\ell_1$}, the position of the center of mass, in the fixed frame, is 
given by
\be
\vect r_{{cm}}=-\ell_{cm} ( \sin  \eta,-\cos \eta \sin \theta,\cos \eta \cos \theta) \equiv -\ell_{cm} \bm u_3
\label{u3}
\ee
With such definition of the fibre direction $\bm u_3$, for small angles $\bm{u}_3 \approx (0,0,1)$.
We note that, for small angles, the two variables $(-\eta, \theta)$ are proportional {(via the distance $\ell_c\simeq0.79$\,m from $P_1$ to the TM center, {see fig.\,(\ref{angoli_cinesi}a))}}  to the  quantities $(x,y)$ measured by the GRS.

We now define  $\bm u_1$ and $\bm u_2$ in such a way that, for small angles, they tend  to $\bm u_x=(1,0,0)$ and $\bm u_y=(0,1,0)$ respectively:

\vspace{.5cm}
\hspace{-.5cm}
\resizebox{.99\textwidth}{!}{
  $
\bm{u}_1'=\dfrac{d \bm u_3}{d \eta}=(\cos \eta ,\sin \eta  \sin \theta ,-\sin \eta \cos \theta),\hspace{0.2cm}
\bm{u}_2'=-\dfrac{1}{\cos \eta}\dfrac{d \bm u_3}{d \theta}=(0,\cos \theta ,\sin \theta).
$
}
\vspace{.2cm}

The minus sign in $\bm u_2'$ is needed to allow the two frames to coincide ($\bm u_3 = \bm u_z$ etc.) for vanishing rotation (and oscillation)  angles.
 Finally, we must allow the TM to rotate around $\bm u_3$ of an angle $\phi$, therefore we redefine $\bm u_1$ and $\bm u_2$  as

\begin{eqnarray}
\bm{u}_1 &=&\bm{u}_1' \cos \phi+\bm{u}_2' \sin \phi \\
\bm{u}_2 &=&-\bm{u}_1' \sin \phi+\bm{u}_2' \cos \phi
\label{u1u2}
\end{eqnarray}

 Equations (\ref{u3},\ref{u1u2}) allow us to assemble the explicit expression for the unit  vectors $\bm{u}_1,\bm{u}_2,\bm{u}_3$. 
 The orthogonal matrix  ${\bm R}$, whose rows are the unit vectors, performs the transformation from one frame to the other:

\be
\hspace{-.1cm}
\resizebox{.93\textwidth}{!}{
$\bm{R}=
\begin{pmatrix}
\cos \eta  \cos \phi &\sin \eta \sin \theta  \cos \phi +\cos \theta \sin \phi&\sin \theta \sin \phi-\sin \eta  \cos \theta \cos \phi\\
-\cos \eta \sin \phi&\cos \theta \cos \phi -\sin \eta \sin \theta  \sin \phi&\sin \eta \cos \theta \sin \phi +\sin \theta \cos \phi\\
\sin  \eta&-\cos \eta \sin \theta&\cos \eta \cos \theta
\label{matriceR}
\end{pmatrix}
$}
\ee
Transformation of an {arbitrary} vector $\vect v$ from one frame to the other frame is achieved by

\be
\left(
\begin{array}{c} 
v_1 \\ v_2 \\ v_3 
\end{array} 
\right) = 
\bm{R} \times 
\left( \begin{array}{c} v_x \\ v_y \\ v_z \end{array} \right); \hspace{1cm}
\left(
\begin{array}{c} 
v_x \\ v_y \\ v_z 
\end{array} 
\right) = 
\bm{R}^{-1} \times 
\left( \begin{array}{c} v_1 \\ v_2 \\ v_3 \end{array} \right)
\label{rotaz}
\ee

\subsection{Angular velocities}
To express the kinetic energy associated to rotation around the centre of mass, we need the inertia matrix and the angular velocity vector, both computed in the moving frame.\\
The components of the angular velocity in the fixed frame are evaluated as follows:
The $\theta$ angle lays on the $yz$-plane, therefore $\dot{\bm{\theta}}$ is parallel to the $x$-axis: $(\dot \theta,0,0)$. The velocity $\dot{\bm\eta}$ has only components along $y$ and $z$: $(0, \dot \eta \cos\theta, \dot \eta \sin \theta)$. Finally, being $\phi$ a rotation around $\bm{u}_3$, we have $\dot{\bm\phi}=\dot \phi \,\bm{u}_3$. Then:
\begin{eqnarray*}
(\omega_x,\omega_y,\omega_z)& 
   {= (\dot \theta-\dot \phi  \sin \eta, \  \dot \eta \cos \theta+ \dot \phi \cos \eta\sin \theta  , \  \dot \eta\sin \theta  -\dot \phi \cos \eta \cos \theta  )}.
\end{eqnarray*}
From this,  using \eref{rotaz}, we find the components of the angular velocity of the body in the moving frame:
\begin{eqnarray*}
(\omega_1,\omega_2,\omega_3)& 
   {= (\dot \eta  \sin \phi + \dot \theta \cos \eta   \cos \phi, \ \dot \eta  \cos \phi -  \dot \theta\cos \eta \sin \phi  , \  \dot \phi+\dot  \theta \sin\eta)}
\end{eqnarray*}

\subsection{Pendulum motion} \label{tretre}
In general, the motion of a pendulum consists of a combination of three different kinds of oscillations, that are colloquialy indicated as torsion, swinging and bouncing.

\emph{Torsion}:
The torsional oscillation is the one that takes place at the lowest frequency, and is therefore the one of interest where weak restoring forces are desired. Its  resonant frequency is given by $\omega_t =\sqrt{\kappa_t/ I}$, where  $\kappa_t = S A^2 /2\pi  \ell_o'$ : $S$ is the shear modulus and $A$ the cross-sectional area  of the fibre; $I$ is the moment of inertia of the TM with respect to the vertical axis (see below). 

\emph{Swinging}:
The swinging pendulum motion, oscillating in a plane containing the $z$ axis, is, in principle, degenerate around  two orthogonal directions. We shall see that, in our case, this is not necessarily the case.  Beside the obvious simple pendulum, at a frequency $\omega_{sw} = \sqrt{g/\ell_{cm}}$ where the fibre bends at the upper end $P_1$, one would also expect a composite pendulum where the TM oscillates around the lower fibre end $P_2$, at a frequency $\omega_{sw2}= \sqrt{mg \ell_1/(I+m \ell _1^2)}$: however this is not found in our data, proving that the fibre does not bend at the point where it connects with the TM.

\emph{Bouncing}:
{The fibre is not rigid: its longitudinal vibrations can be modeled introducing a spring constant $\kappa_b$ 
(see below) and a function $\delta(t)$ to describe the elongation}. 
{The weight of the TM causes a small static displacement of the  equilibrium position, by a quantity equals to $m g/\kappa_b$, therefore we redefine the  fibre length $\ell_0$ = $\ell_0'+m g/\kappa_b$}.
The eigenfrequencies $\omega_n$ of longitudinal waves in the fibre (here considered as a thin rod) are 
{the roots of the equation:}
\be
\tan \frac{\omega_n \ell_0}{c}=\frac{E A}{m \omega_n c}
\label{bounce}
\ee
where $E$ is the Young modulus, $\rho$ is the density of the fibre  and $c=\sqrt{E/\rho}$ is the speed of propagation.
In our case, the first eigenfrequency is found at  $\omega_1 /2 \pi =8.8$ Hz, i.e. $ \omega_1 \ell_0/c \ll 1$,  so that, from eq.\eref{bounce}  $\omega_1^2= E A/m \ell_0$.
All others resonances  ($n>1$) are at frequencies higher than {1 kHz}, well above the range of our interest: we can then treat the rod as a spring with constant $\kappa_b = \sqrt{E A/\ell_0}$. 
In the pendulum here considered we have $ \omega_t /2\pi = 2.2$ mHz, $ \omega_{sw} /2\pi = 0.54$Hz, $\omega_b /2\pi = 8.8$ Hz. The transversal vibrations of the fibre (`violin modes') are also found at frequencies above 100 Hz.

\section{ Lagrangian and equations of motion for the pendulum}
\subsection{The simple, ideal case}

The translational kinetic energy of the TM is

\be
K_{trasl}= \frac{1}{2} m \vert \dot{\vect r}_{cm}\vert^2 \hspace{1cm}\mbox{where}\ \vect r_{cm}=-(\ell_{cm}+\delta(t))\bm{u}_3.
\ee
If the TM is ``accurately'' suspended ({that is, the fibre direction coincides with a symmetry axis})  the inertia matrix is diagonal and the rotational {kinetic} energy is
%

\[
K_{rot}=\frac{1}{2}(I_{11} \omega_1^2+I_{22} \omega_2^2+I_{33} \omega_3^2).
\]
We define the moments of inertia as referred to the suspension point $P_1$ of the fibre:

\[
I_{11}'=I_{11}+m \ell_{cm}^2; \hspace{1cm}I_{22}'=I_{22}+m \ell_{cm}^2.
\]
 In this simple case, the Lagrangian for small oscillations is:

\[
L=\frac{I_{11}'}{2}\dot \theta^2+\frac{I_{22}'}{2}  \dot \eta^2+\frac{I_{33}}{2} \dot \phi^2 +\frac{1}{2}m\dot \delta^2-\frac{1}{2} m g \ell_{cm} (\eta^2+ \theta^2)-\frac{1}{2} \kappa_t \phi^2-\frac{1}{2}\kappa_b\delta^2
\]
and the equations of motion  are: 

\be
\begin{array}{l}\nonumber
I_{11}'\ddot \theta+m g \ell_{cm}\theta=0, \\[5pt]
I_{22}' \ddot \eta+m g \ell_{cm}\eta=0,\\[5pt]
I_{33} \ddot \phi+\kappa_t \phi=0,\\[5pt]
m \ddot \delta+\kappa_b \delta=0.
\end{array}
\ee
We get {four} uncoupled oscillators with {four} different eigenfrequencies. {The first two equations show that the swinging modes are degenerate if  $I_{11}'=I_{22}'$}.
The third equation shows that no other frequency should appear in the $\phi$ spectrum beside $\omega_t =\sqrt{\kappa_t/I_{33}}$: however, this is not what we  experimentally observe.

\subsection{A more realistic, imperfect Test Mass}
In general,  the TM might not be perfectly assembled and, in presence of misalignments,  the center of mass is not aligned with the fibre direction. Therefore, introducing three misalignment angles $\theta_0,\eta_0,\phi_0$, eq.\eref{u3} becomes

\[
 \vect r_{cm}=-(\ell_0+\delta(t))\bm{u}_3+\ell_1( -\sin  \eta_0 \bm u_1 +\cos \eta_0 \sin \theta_0 \bm u_2-\cos \eta_0 \cos \theta_0 \bm u_3)
\]
Moreover, the symmetry axis of the TM too will be misaligned with the fibre. We  then transform from $I$ (which is diagonal) to  {$\tilde{I}$},
according to

\[
\tilde{\bm{I}}=\bm{R}_0^T \times \bm{I} \times \bm{R}_0
\]
with $\bm{R}_0$ given as in \eref{matriceR} with the appropriate angles $\theta_0,\eta_0,\phi_0$. If such angles are small  we find, approximating {to first order}:

\be
\tilde{\bm{I}}=\left(
\begin{array}{ccc}
 I_{11} & (I_{11}-I_{22}) \phi_0 & -(I_{11}-I_{33}) \eta_0 \\
 (I_{11}-I_{22}) \phi_0 & I_{22} & (I_{22}-I_{33}) \theta_0 \\
 -(I_{11}-I_{33}) \eta_0 & (I_{22}-I_{33}) \theta_0 & I_{33}
\end{array}
\right)
\label{Itilde}
\ee
In this case, the rotational kinetic energy becomes

\[
K_{rot}=\frac{1}{2}\sum \tilde{I}_{ij}\, \omega_i\omega_j.
\]
and the translational kinetic energy is

\[
K_{trasl}=\frac{1}{2} m \left[\ell_{cm}^2(\dot \eta^2+\dot \theta^2)+\dot\delta^2+ 2\ell_1\ell_{cm}\dot\phi(\theta_0\dot\eta-\eta_0\dot\theta)-2\dot\delta(\eta_0\dot\eta+\theta_0\dot\theta)\right].
\]
In the same approximation, the potential energy, including the torsion term, is

\[
U=\frac{1}{2} m g\left[ \ell_0 (\eta^2+ \theta^2)+\ell_1((\eta_0+\eta)^2+(\theta_0+\theta)^2)\right]+\frac{1}{2} \kappa_t(\phi-\phi_0)^2+{\frac{1}{2}\kappa_b \delta^2}.
\]

A further semplification occurs if $I_{11}=I_{22}$: this is certainly verified in our experimental case to better than 1\%.   Then, in \eref{Itilde} we have $\tilde{I}_{12}=\tilde{I}_{21}= 0$. { We now perform a translation to new variables, measured from the equilibrium position:  $\theta_{eq}=-\theta_0 \ell_1/\ell_{cm},\eta_{eq}=-\eta_0 \ell_1/\ell_{cm},\phi_{eq}=\phi_0$. The  simplified Lagrangian leads to the following equations of motion}:

\be
\begin{array}{l}
I_{11}'\ddot \theta+m (g \ell_{cm} \theta-\ell_1 \theta_0 \ddot\delta)- \eta _0 q_\theta \ddot \phi=0,\\[5pt]
I_{11}' \ddot \eta +m (g \ell_{cm} \eta-\ell_1\eta_0\ddot\delta) + \theta _0 q_\eta \ddot \phi=0,\\[5pt]
I_{33} \ddot \phi+\kappa_t \phi- \eta _0 q_\theta\ddot\theta+\theta _0 q_\eta\ddot\eta=0,\\[5pt]
m \ddot \delta+\kappa_b \delta-m \ell_1(\eta_0\ddot\eta+\theta_0\ddot\theta)=0,
\label{eq_fin}
\end{array}
\ee
where

\[
q_\theta=I_{11}-I_{33}+\frac{\ell_1}{\ell_{cm}}(I_{33}+m \ell_{cm}^2),\hspace{1cm}q_\eta=I_{11}-I_{33}+m\ell_{cm}\ell_1.
\]
Being $I_{33}\ll I_{11}\ll m \ell_{cm}\ell_1$ we assume, from now on,  $q_\theta\approx q_\eta\approx m \ell_{cm}\ell_1$. Note that the introduction of misalignments generates coupling among all variables.
The third of eq.\eref{eq_fin} shows that the coupling between torsion and swinging arises from both $\theta_0$ and $\eta_0$ angles of misalignment.  This can explain the last point in the list of  unexpected features, i.e. the modulation of the torsion motion at the swinging frequency. Finally, the fourth equation contains $\theta$ and $\eta$: this explain why the doublet is also present in the $z$ spectrum (second point of the list of unexpected features).
 
We solve the system \eref{eq_fin} in the frequency domain by assuming  sinusoidal time dependence for all variables ($\theta (t) = \tilde\theta\, {\rm e}^{i \omega t} $ and so on)

\be
\begin{array}{l}
\tilde\theta(m g \ell_{cm} - \omega^2 I_{11}')+\omega^2(  q\,\eta_0 \tilde\phi  +\theta_0 m \ell_1  \tilde\delta ) =0,\\[5pt]
\tilde\eta (m g \ell_{cm} - \omega^2  I_{11}') - \omega^2 (  q\, \theta_0 \tilde\phi-\eta_0 m \ell_1\tilde\delta  )=0,\\[5pt]
\tilde\phi (k_t - \omega^2 I_{33} )+\omega^2 q ( \eta_0  \tilde\theta-  \theta_0 \tilde\eta  )=0,\\[5pt]
\tilde\delta (k_b -  \omega^2 m) + m\ell_1 \omega^2 (\eta_0 \tilde \eta + \theta_0 \tilde\theta)=0.
\label{eqfreq}
\end{array}
\ee

the normal mode frequencies are\\

\resizebox{.95\textwidth}{!}{
\hspace{-1cm}
    $\begin{array}{l}
     \vspace{0.3cm}
     \omega_{0,2}^2=\dfrac{I_{11}' k_t + I_{33} m g \ell_{cm} \pm \sqrt{  I_{11}'^2 k_t^2 +m g \ell_{cm} \left( I_{33}^2 m g \ell_{cm}- 2 k_t (I_{11}' I_{33}  - 2 q^2 (\eta_0^2 + \theta_0^2))\right)}}{2 (I_{11}' I_{33} - q^2 (\eta_0^2 + \theta_0^2))}\\
     \omega_{1,3}^2=\dfrac{I_{11}' k_b + m^2 g l_{cm}  \pm \sqrt{I_{11}'^2 k_b^2   +  m^2 g l_{cm}  \left( m^2 g l_{cm} - 2 k_b ( I_{11}' -2 \ell_1^2 (\eta_0^2 + \theta_0^2))\right)}}{ 2 m (I_{11}'  -  \ell_1^2 (\eta_0^2 + \theta_0^2))}
     \end{array}$%
  }

\vspace{0.5cm}
\noindent
$\omega_0$ corresponds to the torsion frequency, $\omega_1$ and $\omega_2$ are the swinging modes and $\omega_3$ is the bouncing mode. 

\subsection{Evaluation of the misalignment from our data} \label{numeri}
We now  insert numerical values\footnote{Numerical values of the parameters for  our apparatus, used in sect. \ref{numeri}: $\kappa_b=299.6$\,kg m$^2$ s$^{-2}$, $\kappa_t= 6.77\cdot 10^{-9}$\,kg m$^2$ s$^{-2}$,  $I_{11}=I_{22} =2.76\cdot 10^{-4}$\,kg m$^2$; $m=0.106$\,kg, $\ell_0$=0.65\,m; $\ell_1$=0.11\,m, $q=7.2\cdot 10^{-3}$\,kg m$^2$.} for our pendulum, and proceed to evaluate the two misalignment angles: 
we expand the above relations in series of $\theta_0^2+\eta_0^2$ , obtaining
\begin{equation}
\begin{array}{l}
\nu_{0}\approx 0.0022-5.905 \cdot 10^{-7}\,(\theta_0^2+\eta_0^2)+\dots \; {\rm Hz}\\[5pt]
\nu_{1}\approx 0.5582 -2.322\cdot 10^{-5}\,(\theta_0^2+\eta_0^2)+\dots \; {\rm Hz}\\[5pt]
\nu_{2}\approx 0.5582 + 10.156\, (\theta_0^2+\eta_0^2)+\dots \; {\rm Hz}\\[5pt]
\nu_{3}\approx 8.811 + 0.0913\, (\theta_0^2+\eta_0^2)+\dots \; {\rm Hz}
\end{array}
\end{equation}
The frequencies of the torsion and bouncing modes are virtually unaffected.
On the contrary, the misalignment generates a splitting of the swinging frequency. The separation between the two frequencies (14 mHz in our case) is proportional to the square of the misalignment angles:
\[
\nu_{2}-\nu_{1}= 10.156\, (\theta_0^2+\eta_0^2)=0.014 \; {\rm Hz}
\]
that gives $\sqrt{\theta_0^2+\eta_0^2}= 37 \ mrad$.
{As mentioned in sect.\ref{sect2}, the frequencies in the doublet decouple under rotation on the $xy$-plane  (or  ($-\eta,\theta$)-plane). Indeed,  we can  apply  such rotation: 
\begin{eqnarray*}
\tilde\theta&=&\tilde\theta' \cos \alpha+\tilde\eta'\sin \alpha\\
\tilde\eta&=&-\tilde\theta' \sin \alpha+\tilde\eta'\cos \alpha
\end{eqnarray*}
and  re-write the equations (\ref{eqfreq}) (discarding the negligible  bouncing terms). By eliminating  $\tilde\phi$  between the first two we get:}
\[
(m g \ell_{cm}-I_{11}'\omega^2)(\tilde\theta' ( \theta_0 \cos \alpha- \eta_0\sin\alpha)+ \tilde\eta' (\theta_0 \sin \alpha+\eta_0\cos\alpha))=0.
\]

With two particular values of $\alpha$, the physical pendulum mode (which corresponds, with good approximation, to $\nu_1$)  can be isolated along the $\eta'$ or $\theta'$ direction. The values are, respectively
\be
\alpha_{1}=\arctan \frac{\theta_0}{\eta_0},  \mbox{    and its complement   } \alpha_{2}=-\arctan \frac{\eta_0}{\theta_0}.
\label{al1al2}
\ee
We found from the data  $\alpha_2\approx13^\circ$: this and the previous relation imply $\theta_0
\simeq \pm 36$ mrad, $\eta_0 \simeq \mp 8.4$ mrad.
\section{Comparison with the data and  their interpretation}\label{data}
As mentioned above, the motion in the $(x,y)$ or, equivalently, $(-\eta,\theta)$ plane, describes Lissajous figures.We have shown that this unexpected feature can be generated by a misalignment of the TM of $\approx 2^\circ$ and the inclination of the envelope of the trajectory is strictly connected to the misalignment angles, (eq.\,\ref{al1al2}).
After a rotation of $\alpha_1$ (or $\alpha_2$), each of the two signals appears on only one axis. 

\begin{figure}[h!]
\hspace{-1cm}
\includegraphics[width=0.49\columnwidth]{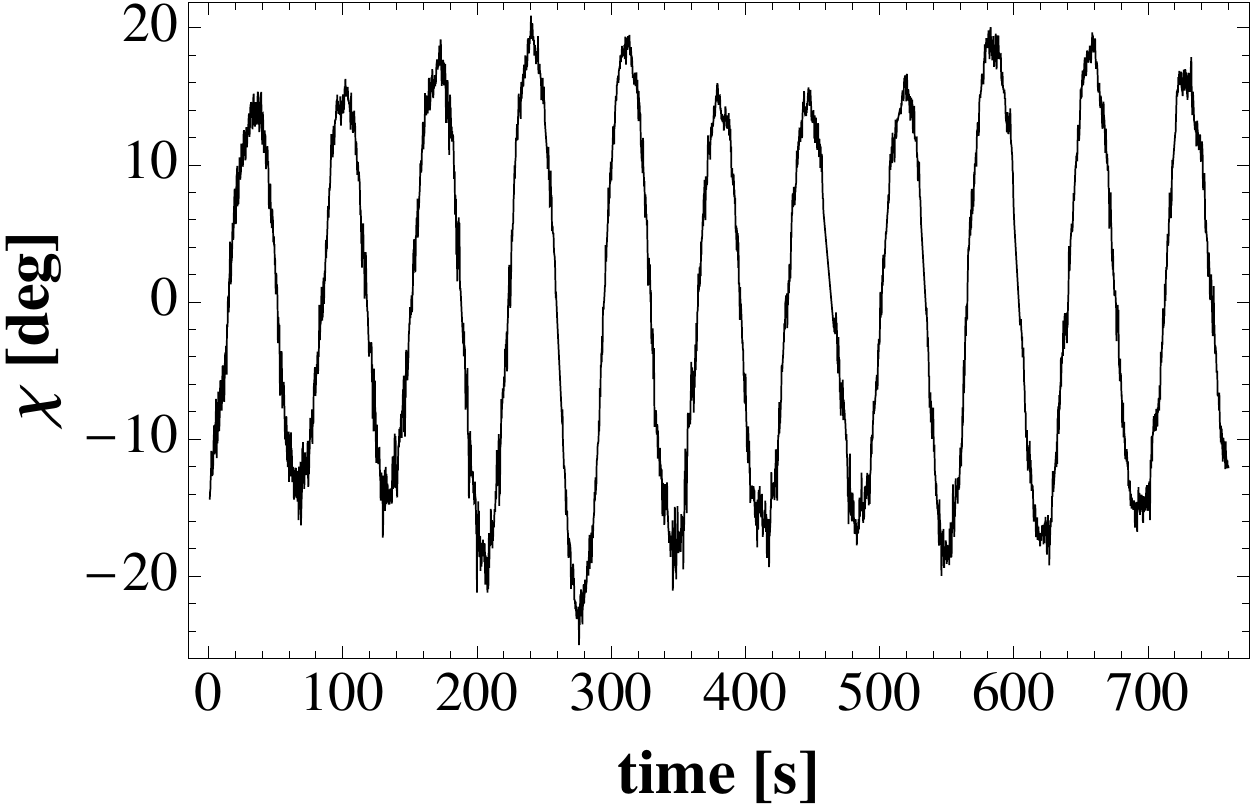}
\includegraphics[width=0.49\columnwidth]{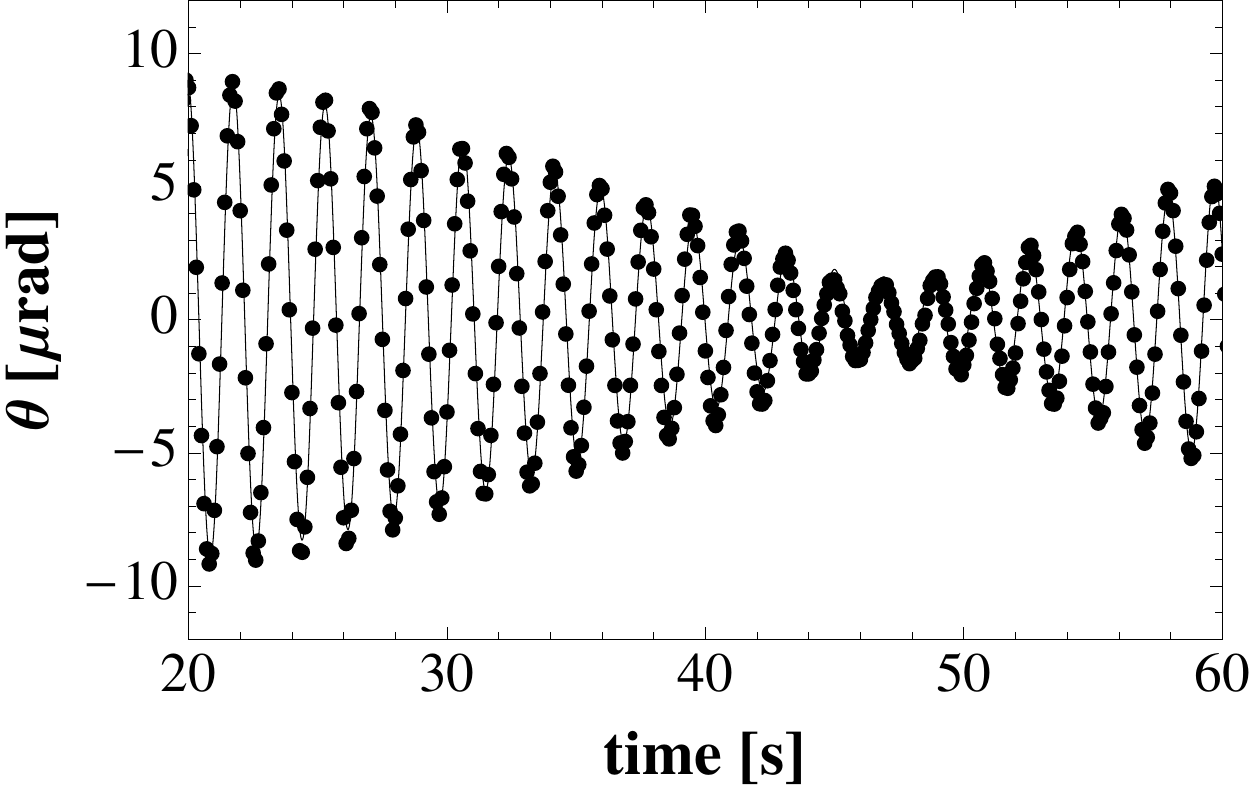}
\includegraphics[width=0.49\columnwidth]{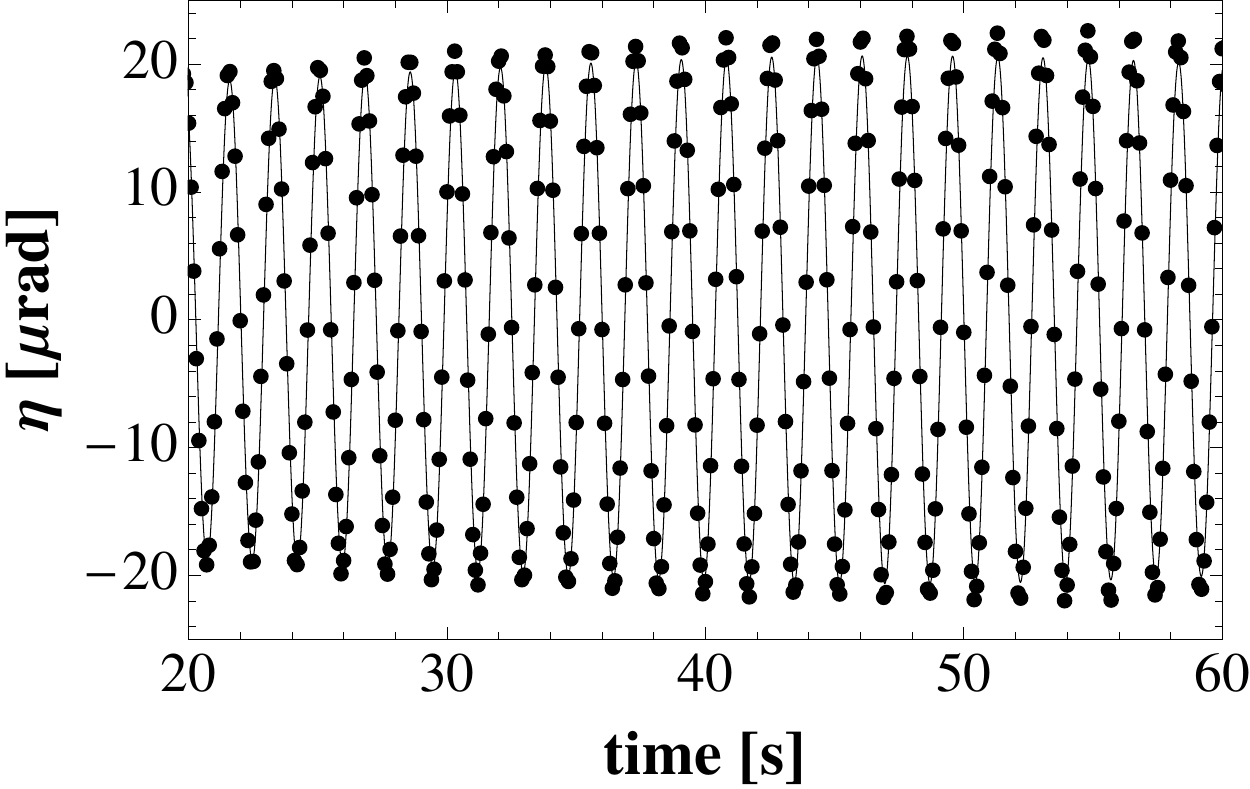}
\includegraphics[width=0.49\columnwidth]{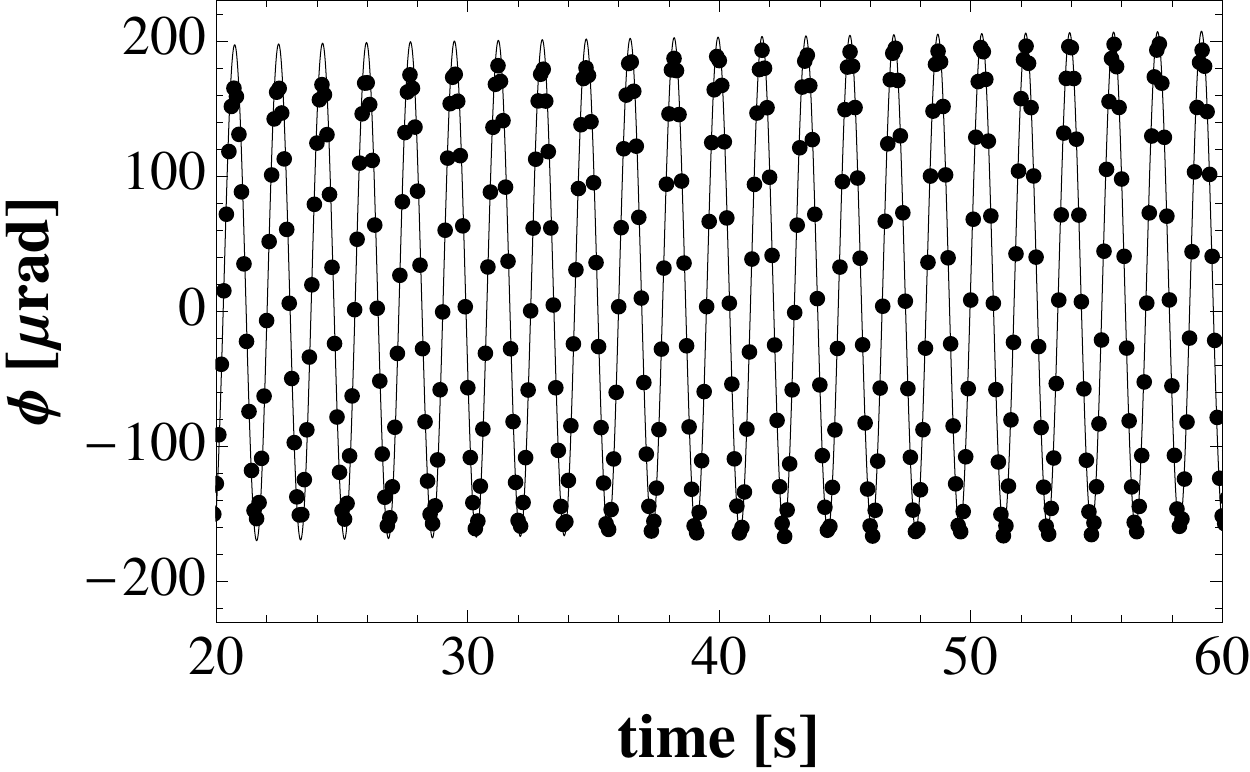}
\caption{{Top left:  libration angle of the osculating ellipse, derived by pendulum data. Note the periodicity at  T=72 s, the inverse of the beat frequency.
Other panels: fits of  $\theta(t)$ (top right), $\eta(t)$ (bottom left) and $\phi(t)$ (bottom right). Dots: measured values, line: best fit.  For clarity, only a 40 s stretch  is shown, but the quality of the fit is constant over the entire 500 s span.}}
\label{fit_e_libra}
\end{figure}

The nature of the motion in the $xy$-plane can be assessed by computing the `libration angle' $\chi$ of the osculating ellipse by means of the formula \cite{bp}

$$ \tan 2\chi= 2 \frac{\dot x \dot y +  \omega_1 \omega_2 x y}{\dot y^2 +  \omega_2^2 y^2-\dot x^2 -  \omega_1^2 x^2} $$
{where $x(t), y(t)$ are the data and their derivative are computed by spline interpolation.} In the {top left panel} of Fig.\,\ref{fit_e_libra} the libration angle of the osculating ellipse, obtained from pendulum data, is plotted on a time span long enough to  show its evolution: it is clearly noticeable {that this ellipse librates at the beat frequency of the doublet, i.e.  14 mHz.}

In order to verify our model, we have tried to reproduce the observed data  with a least-square fit procedure: we selected a sample of data over 500s (roughly a single torsional period) and fitted the three observed outputs ($\phi(t),  x(t) =- \ell_c \cdot \eta(t)$  and $y(t) = \ell_c \cdot \theta(t)$) with a sum of three sine waves, leaving the initial conditions, the amplitudes and the misalignment angles as fitting parameters.
We did not fit the $\delta$ coordinate because the  bouncing frequency falls above the Nyquist frequency and we can only observe the aliased signal. 

We find a very good agreement for $\theta$ and $\eta$, but, regarding the modulation of $\phi$ at the swinging modes, the component at $\nu_2$ results overestimated with respect to the observed data.
We remark that our analysis is still incomplete, having neglected several possibly relevant cause of discrepancies, e.g.:

\begin{itemize}
\item errors or drifts in the calibrations that convert the data from volt to meters,
\item possible misalignments of the GRS with respect to the local gravity direction,
\item damping of the modes,
\item  {transverse} elasticity of the fibre (here modeled as a rod).
\end{itemize}

 {As an example, regarding the first of the above items, the fit quality can be substantially improved by allowing small differences in the electronic gains of the various channels (we recall that the same channels are used in different combinations to provide the various DOFs). }


{We observe that both the above found misalignment angles and the proposed gain unbalances have quite large values.} For this reason, we mention another mechanism that could also originate the frequency splitting of the swinging signal:
consider an anisotropic external force with components in the $xy$-plane only:

\[
U_{ext}=\frac{1}{2}(k_1 x^2+k_2 y^2)\simeq \frac{1}{2} \ell_{cm}^2 (k_1 \eta^2+k_2 \theta^2).
\]
 Combined with the gravity restoring force, it produces two different frequencies in the two horizontal directions.

\[
\hspace{-2cm}
\frac{\ell_{cm}(m g+k_1 \ell_{cm})}{I_{11}'}=\omega^2_{1} \hspace{0.5cm}\mbox{and} \hspace{0.5cm}\frac{\ell_{cm}(m g+k_2 \ell_{cm})}{I_{11}'}=\omega^2_{2}. 
\]
From these, using measured values we find $k_1=0.0144$ \ kg s$^{-2}$ and $k_2=-0.0469$ \ kg s$^{-2}$, i.e. two values that are again too large not to raise suspicion.

Although this field appears as ``added ad hoc" to explain the data, it does have physical bases to rest on, as several possible sources of anisotropic coupling exist. 
We mention a few among the more obvious:  residual magnetization of the TM coupling to the external field,  electric patch effects, density inhomogeneities and machining defects in the TM, asymmetric fastening or bending of the fibre \cite{lp}.  It is likely that a combination of some of the above mentioned mechanisms (e.g. misalignment and external field) is the real cause of the observed anomalies.
%

\section{Conclusions}\label{fine}

{We have developed a model that explains, on pure geometrical basis, several odd features found in our pendulum data.  A detailed analysis of the geometry of the Test Mass and its suspension in a non-ideal case leads to coupled equations of motion that simply explain  the unexpected couplings. We have shown that a detailed mechanical model of the torsion pendulum, where geometrical imperfections are taken into account, can explain two unexpected features of our data: the modulation of the torsion signal at the natural frequency of  the swinging motion and the splitting of the swinging resonance.  The misalignment angles needed to fit the data and interpret all their features are however larger than expected on experimental basis. We have also shown that an additional, anisotropic  elastic field, whose origin remains to be determined, can also account for the second of these features;  the required anisotropy is again quite large. The geometrical model here developed may help in dealing with anomalies of similar kind, that can also be observed in different experimental set-ups.}

This work was undertaken as an intermediate step toward the development of a more complex model for a double pendulum with two soft degrees of freedom \cite{8dof}:  we believe it can form a useful basis to a better understanding of both simple and composite torsion pendulums.

\section*{Acknowledgments}
 We thank L. Di Fiore, R. De Rosa and F. Garufi for useful discussions. Continued advice and support from the Trento LISA-Pathfinder group, and in particular from W.J.Weber,  is appreciated. Work supported by MIUR (grant PRIN 2008) and INFN.


\end{document}